# ARIS

An open source platform for developing mobile learning experiences

by David J. Gagnon

# Abstract

Inspired by mobile, Internet enabled computing and the maturing field of educational game design, the ARIS project has designed an open source tool for rapidly producing locative, interactive, narrative-centric, educational experiences. In addition to the software, the project contributes a global community of active designers and a growing set of compelling mechanics for learners in such designs.

# Acknowledgments

Since I began working on this project in 2008, I've been surprised that the ideas contained never failed to attract passionate people. By far, the most rewarding part of this work has been the community of fellow explorers, who have remained invested from the first day until the present, playing roles big or small, many who have worked without payment.

The members of the ARIS team, past and present*:

| | | |
|---|---|---|
| Chris Blakesley* | Seann Dikkers* | Kevin Harris |
| Yoonsin Oh | Ryan Martinez | Peter Debbink |
| Matt Gaydos | Jim Matthews* | John Martin |
| Chris Holden* | Mark Wagler | Thomas Henage |
| Ron Cramer | Brad Leege | Dee Johnson* |
| Brian Deith | Daniel Gosch | Muhammet Demirbilek |
| William Buck | | |

A most profound respect and gratitude is extended to Dr. Kurt Squire who has directed the path of this project from its conception and who has provided overwhelmingly generous support in terms of funding, personal encouragement and scholastic perspective.

This project also would not have happened without the support of Carole Turner and Chris Lupton at DoIT Academic Technology who have made space for its growth and provided the initial seed funding to create an authoring tool for students.

A final thank you goes to my dear wife, Sarah, who has given considerable amounts of herself discussing and affirming this work, continually providing wisdom about the things that are important and those that are not.

# Table of Contents

# Introduction

ARIS, Short for Augmented Reality and Interactive Storytelling, is an authoring tool as well as an iPhone application that work together to create mobile, locative, narrative-centric, interactive experiences. As a design response to a number of theories and examples from various disciplines of learning science, curriculum studies, media studies, contemporary social media and game design, it serves a complex continuum of user-designers including artists, teachers, students, administrators and researchers who each use it in a different capacity ranging from a rapid prototyping tool for interactive stories to a mobile scientific data collection tool.

As the ARIS software has matured and become increasingly useful, much of the project's effort has become directed toward the design of communication tools, collaborative processes and community building efforts for each of the various audiences in effort to distribute the design, development, and dissemination of ARIS and it's content across a large and hopefully stable group of independent collaborators.

In this manuscript I wish to highlight specific topics from the education and design literature that make a case for a specific set of attributes that informed ARIS's development. Then attention will turn to the project narrative from a first person perspective, highlighting lessons in community building, agile development processes and iterative, user centric design methods. In the final section, ARIS adoption statistics and outside project collaborations will be discussed.

# Inspiring Thoughts from The Literature

## The Importance of Media

The value of exploring new media and technology, such as mobile devices and video games, in pursuit of learning about better curricular design, is of grave importance. Not only are new medias such as Internet enabled mobile devices, social networking, micro-blogging and video sharing seeing steady adoption increases among students (Smith & Caruso, 2010) but the nature of those medias, possibly more than the content, inform the way those students and society at large think (McLuhan & Lapham, 1994). Merlin Donald (1993) describes how the specific anatomy of the brain has changed by the historical adoption of new communication methods. Along similar lines, Donald Norman (1994) submits that the technology humans interact with influence the actual ways cognition takes place, moving increasing amounts outside of the mind, embedding knowledge in the artifacts we design and surround ourselves with. It seems that society, anatomy and cognition are deeply connected in and around media.

Narrowing the discussion of media to educational design, we see the discussion is vibrant here as well. Kosma (1991,1994a,1994b) engages directly in the debate about media and it's role in education, claiming that it has the power to influence the structure, formation and modification of mental models and therefore should be considered a primary concern of educational designers.

## Cultural Phenomena as a Result of New Media

There is also value in understanding the cultural practices that have been enabled by new media and how they have shifted the ways that people use and produce knowledge. In recent years, the freely-available host of so-called Web 2.0 communication tools have highlighted examples of group learning and thinking phenomena. Built on the theory that cognition itself can be distributed across groups, tools and environments (Hollan, Hutchins & Kirsh, 2000), these tools are enabling a method and even a motivation for individuals to collaborate and produce emergent knowledge (Cress & Kimmerle, 2007). The high volume, low latency nature of electronic communication allow knowledge communities (Paavola, Lipponen & Hakkarainen, 2004), communities of practice (Lave & Wenger, 1991) and place-less organizations (Nardi, 2007) to be formed and sustained in a way that could not happen otherwise. Because the cost of reproducing information is diminishing and digitized artifacts can be given without given away (Wiley, 2010), phenomena such as open source software and open educational resources are increasingly feasible options for producing globally useful assets that retain customizability to local needs.

## Open Curriculum and Democratic Education

While commercial textbooks and peer-reviewed journals are increasing in cost each year, causing some libraries to actually reduce their journal catalogs (Dingley, 2006), the use of open access alternatives is steadily growing. For example, one of the larger open textbook projects, Connexions[1], currently provides over 4,000 textbooks for modification and use. Brainchild of Richard G. Baraniuk, an engineering professor at Rice University, Connexions receives 850,000 unique users a month, with more than 50 percent of the traffic originating from outside the United States (Baraniuk 2008). Other institutions are participating in the Open Educational Resource (OER) community as well. UC Berkley, MIT, Utah State, Yale, Standford and many others are beginning to make lectures and course notes freely available. Some allow modification and redistribution as well.

---

[1]Other open content editing and distribution networks exist such as WikiEducatior and the OER Commons

While there are hurdles in place for OER, such as the view expressed by some authors that publishing in an open access publication reduces their chance of receiving grants or promotion based on that work (Swan and Brown, 2004), open access with the ability to customize may actually lead to a more democratic process of curriculum production and selection. Michael Apple (2000) states that commercial textbooks are a knowledge artifact representing the agenda of hegemonic political interests. This attribute of commercial curriculum is stark contrast to OER's key tenant that education can be improved by harnessing the collective wisdom of a community of practice and reflection (Liyoshi and Kumar, 2008). Gabriella Coleman (2004) drives the point home, stating that free and open source software, which provided the original inspiration to, and shares many attributes with OER (OECD, 2007), has the power to be politically agnostic.

## Games as a Promising Educational Media

While the promise of OER and democratic production of curriculum is compelling, the advantages of new media go far beyond its methods for production and distribution, enabling new forms of interaction and experience that would be otherwise unattainable. For example, the ability for the reader to influence the choices of the main character in a "Choose your adventure" book fundamentally shifts the ways in which the narrative is experienced. The reader has become a participant, co-designing experiences and characters with the original author as they make choices that have consequences. Video games are, by their very nature, built around this kind of interaction and participation.

Video games, therefore, provide a means for designing curriculum that surpasses simple exposure to content but aims to enrich student experience through active participation (Squire 2006, Gee 2004, Dewey 1938). An example of this approach is seen in the engineering video game *Cool It*, developed at the University of Wisconsin to teach a unit in cryogenic engineering (Pfotenhauer 2009). In this game, students solve challenges with virtual lab equipment, seeing the results of their choices and building tacit understandings about material properties and cryogenic principles (Pfotenhauer, In Press). Student activity in this game extends into experimentation and practice, well beyond exposure or memorization. The dominant strategy of the game is not being able to recall a formula but reverse-engineering the mathematical relationships through experimentation and play.

Video games also allow students to take on new identities and abilities beyond their own expertise. In the game *Rock Band* (Harmonix, 2007), non-musician players are put in the "shoes" of rock stars, playing large venues of cheering fans and going on national tours. Even though they are only pressing keys on a plastic guitar or singing out-of-tune into the supplied USB microphone, the game mediates their input, generating musical outputs that only experts would be able to produce. In social games and virtual worlds, this ability to take on the identity other than ones own, even inventing new identities, has a powerful effect on the participants, in some cases even acting as a form of psychotherapy (Turkle, 1995).

Feedback is one of the most significant activities a teacher can engage in order to improve student achievement (Hattie, 2004) but any delay in providing feedback diminishes its value for learning (Hattie, 2004; Bangert-Drowns, Kulik, Kulik & Morgan, 1991). As games are designed systems of cause and effect relationships, they provide a unique opportunity to produce real-time as well as reflective feedback based on player activity (Pfotenhauer, In Press). Schell (2008) even argues that timely and meaningful feedback is a prerequisite for good game design.

Games also provide a means for producing new forms of narrative (Salen and Zimmerman, 2003; Jenkins 2002). Bruner (1991) and Polkinghorne (1988) claim that a primary method of meaning making relies on the narritivization of experience sequences into mental schemata. Not only does thought occur in terms of story but also previously understood narratives are used for creating schema from new experience. (Schank & Abelson, 1977; Schank & Abelson, 1995).

## Mobile as a Promising Technology to Combine Play, Place and Production

Enabled by the MIT-OAR augmented reality activity authoring tool made by Eric Klopher and team at MIT, a number of researchers have probed how location-aware mobile devices may be used to create educational video games (Klopher, 2008). A variety of mobile learning designs such as environment simulation (Klopher & Squire, 2008), design literacy (Matthews, 2010) and scientific literacy and argumentation (Squire & Jan, 2007) have demonstrated promising results. Mobile media and augmented reality has a unique ability to unite the advantages of educational video games with place-based learning (Squire et al., 2007).

Mobile may be particularly suited for creating educational experiences in informal settings, whether it is just-in-time information about how to navigate a public transportation system or downloading guitar chords to play a song from the radio (Gagnon, 2010; Squire, 2009). However, little is known about how youth interact with media in those circumstances (Squire, 2009) or how learning experiences for those contexts should be designed.

## Iterative Development and Research Methods

In addition to theoretical readings, methodological approaches have also provided inspiration for this project. The single most influential work, as short as is it, was the Agile Manifesto (Beck, 2001), co-written by 17 software developers and copied here in it's entirety.

> We are uncovering better ways of developing software by doing it and helping others do it. Through this work we have come to value:
>
> Individuals and interactions over processes and tools
> Working software over comprehensive documentation
> Customer collaboration over contract negotiation
> Responding to change over following a plan
>
> That is, while there is value in the items on the right, we value the items on the left more.

The manifesto is elaborated with a set of 12 principles that involve self-organizing, small teams, frequent delivery, function over documentation, working at a sustainable pace and the value of motivated individuals.

Scrum (Schwaber & Beedle, 2001) is a particular method of agile software development built around the notion of performing work iteratively in units of time called *sprints*. The goal of each sprint, usually 2-4 weeks, is to produce some sort of deliverable and incremental value to the project. For this to work, all the bugs, feature requests and chores that have been identified are placed in a prioritized queue called the *backlog*. This list is reordered, placing the most valuable items at the top, based on the Scrum master's current assessment. At the beginning of a sprint, the developer team makes a prediction as to how many of the items on the top will be completed and then begins working, sticking exactly to the order of the backlog items, which have been frozen for the duration of the sprint. The sprint ends based on the original time box, not if the predicted tasks were performed. Any remaining items are returned the backlog for the next sprint. In this way, the team learns to predict its production rate more and more accurately each sprint, priorities can shift quickly while not confusing the current work and incremental improvement of the product is always the most valued goal.

In the education research world, these incremental and agile software development approaches resonate of the features described in Design Based Research (DBR) (Barab & Squire, 2004; Design-Based Research Collective, 2003). DBR is based on the idea that "Learning, cognition, knowing and context are irreducibly co-constituted and cannot be treated as isolated entities or processes" (Barab and Squire, 2004) and that research must be performed by designing treatments that lead to results within authentic contexts that will hopefully lead to theories that extend past those particular contexts. Like the Agile Manefesto and Scrum method, DBR focuses on tangible deliverables in specific contexts with iterative approaches.

# The ARIS Project as a Design Solution

It would be ridiculous to say that all of the above influences were taken into account upfront and weaved into a single unified design for the purpose of mobile learning research. In contrast, the ARIS project has been a three year process of following design hunches while staying in contact with various audiences of use. It has been full of backtracking and reevaluating, starting code from scratch four times to keep pace with changing goals and technology advances. Outside of Chris Blakesley and myself, the core team has varied wildly, involving dozens of members, some for years and others for only days. But that is not to say that ARIS has had no stable foothold. In retrospect, it is quite clear that a particular collection of education and design scholarship combined with a shared aesthetic vision has provided a foundation on which a living project has been built. In many ways, the current piece of software and community is a design solution to combine all of the influences documented above.

## Inspired by the iPhone and Sam Cooke

In 2007 the iPhone was released and a new excitement about mobile devices was brewing. Photographs and news footage showing the long lines of customers waiting for the new product outside of Apple© stores surprised and tempted us to join the bandwagon. The device was unlike its predecessors in that it elegantly combined a massive screen with a multi-touch interface, an always-on broadband Internet connection, a built in camera, GPS, accelerometers and a host of applications like google maps and mobile safari into a pocket sized package that also functioned as a phone. I was convinced that we had entered a new era of personal computing and that within five years or so everyone would have a device like this.

While brainstorming ideas for an educational game to produce as part of a class activity, I simultaneously remembered an art history course I had attended as an undergraduate and a set of lectures from Craig Werner, author of *A Change Is Gonna Come: Music, Race and the Soul of America* (Werner 1999). The art history course was taught in a massive lecture room of moderately interested students, a projection flickering at the front of the space and background drone of a lecture's voice. Assessment, as it does, reflected the genuine values of the course's instruction, namely the ability for a student to remember the name, author and date of a work when presented with an image. A perfect instantiation of this form of knowledge, flip-cards were a very popular form of study aid among the students.

For Werner's book however, art was taught as a conversation likened to the rowdy call and response sermons found in a gospel church. He explains that artists send out the "call" like a preacher each time they release a song to the radio or a record to the stores. If the community resonates with the message, they send back a thunderous "Amen!" or "Bring it!" by buying the record or attending those shows. Sometimes other artists respond to the call by producing another piece of work, referencing the original. In this way, an artist that knows how to listen for the response will find their voice and update their message, sending out a new call with their next work. A national conversation can be traced through the lyrics, musical "impulses" and resulting popular response of each top 40 hit if the observer is aware of the call and response pattern.

I found this view particularly inspiring as Werner was able to extend this notion of public conversation to the chronologically parallel events of the civil rights movement. This analysis reveals the profound message of songs like Sam Cooke's, *A Change is Gonna Come* (Cook, 1964), allowing listeners to empathize with the complex context in which it was written and the desperate ache of the audience it calls to. Popular music, in this view, is more than simple entertainment. It is a medium for (D)iscourse (Gee, 2004) within a society.

I wanted to do this for Art History. I wanted to understand the call and response of the art world and how it affected the paths of the artists involved and their onlookers. Following the metaphor of conversation, the ad-hoc team who responded to class project pitch began selecting the characters and writing the branching dialog that would compose a game about the Pop Art movement in New York during the late 1950s and 60s.

In effort to construct an interactive narrative out of immutable historical fact, we settled on what we called the "Forrest Gump" method of history. This meant that we would place the player/student in a fictitious historical role, but have their actions causative for known historical events. This was demonstrated in the film when Forrest Gump taught Elvis Presley his famous dance moves or was made into the inventor of the "Shit Happens" slogan. In the game we designed, the student plays an up-and-coming sculptor who is trying to get invited to one of Andy Warhol's infamous studio parties by hobnobbing and helping the various artists of the movement find their muse. The climatic scene involved delivering a Campbell's Soup can to Andy Warhol from Roy Lichtenstein; a can that would later become a centerpiece of Pop Art.

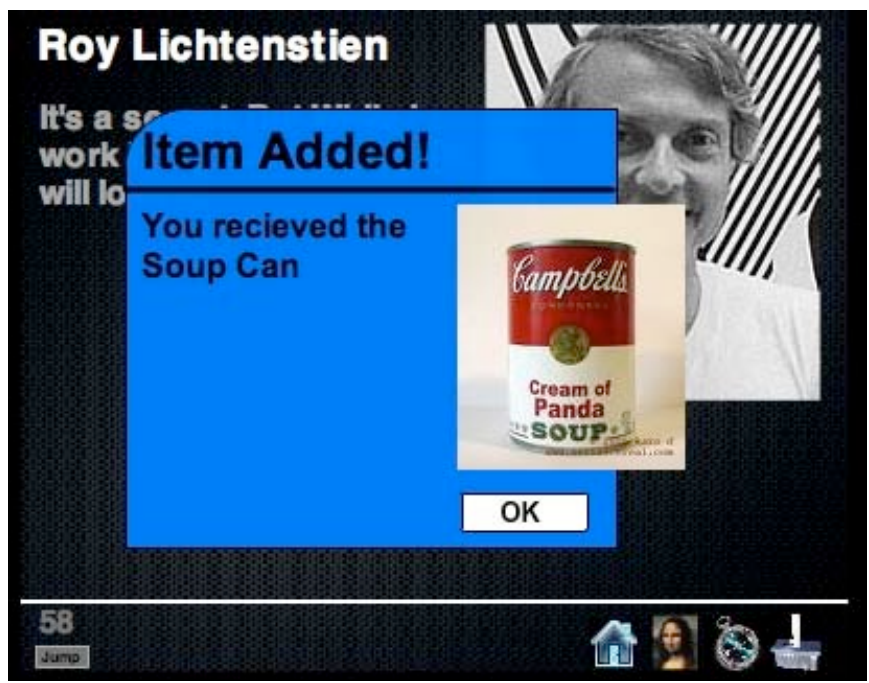

A screenshot from Get In With Warhol, built in ARIS 0.09

The project, "Get In With Warhol" and was built on the newly created ARIS 0.09 platform. Technically speaking, it was a Flash application that ran in a web browser on Microsoft Pocket PC devices and was driven by an XML file that could be edited from Microsoft Excel. The implementation plan was to embed numeric codes into the Milwaukee Museum of Art, which has a sizable Pop Art collection, next to various locations where the virtual artists would reside. By typing in a code, the device would know where you were and what characters were nearby. The prototype was never implemented in the museum, only in a room in the Teacher Education Building for playtesting purposes.

The team should have worked brilliantly; there was an art educator and a few people with film and storytelling backgrounds. Unfortunately, the complexity of authoring interactive narrative combined with the task of translating those ideas into rows and columns in Excel was more complex than anyone expected. Not only did we need to learn more about how to make these kinds of stories but how to create a system to play them that was accessible by non-programmers.

## Discovering the Mechanics of Mobile Activity

Seann Dikkers joined Chris Blakeley and myself as we continued experimenting with mobile learning design and creating the ARIS platform. Starting from teacher needs to find creative materials that teach to specific state standards, the team went through a process of brainstorming game ideas that would be practical and within scope of available technology. We settled on curricula standards relating to US and WI history and headed out to the state capital to see what kind of activities we could dream up. The final game design was called Saving Dr. Hernandez and was a locative puzzle game, in the line of Myst (Brøderbund, 1993), about a time traveling archaeologist who had been lost in the past and needed to be located through following clues that were left in the design and artifacts of the WI State Capital building.

Surprisingly, the Capital building had no WiFi Internet access, so we needed the system to work off-line. Because the game would be played indoors, GPS would not be available for location detection and we needed to come up with other ways of making sure the narrative remained linked to the space the player was in. These challenges, though disappointing at the time, encouraged us to think about the ways that a computer can determine if the player is in a particular location using other means. In the end we used five main play mechanics and two methods for detecting location.

In the first segment of the game, players needed to find a particular historically significant statue in the southeast wing called *The Genius of Wisconsin*. Interestingly, the statue did not appear in any of the printed material available at the Capital or in any easy online searches. The seemingly simple task of finding it required asking one of the guards or guides, who all seemed to know exactly where to go. We realized that in many environments, unaware of our game, staff would respond in particular ways that could be relied on and built into the activity. This was the first mobile game mechanic we discovered. These were activities that require talking to real people in order to gain game-relevant information. We called the mechanic *unknowing actors* and used it in later designs as well.

To prove to the system that the player had correctly identified the space, a set of branching multiple-choice questions would ask about statue details. Questions such as, "Is the Genius a man or a woman?" would quickly reveal if the player had indeed found the statue and was currently in front of it. Obviously this method would not work if players shared information with each other, a flaw of many games in this genera.

The second segment of the game also illustrated a specific mobile mechanic. Now that the player was standing in front of the statue, they were given a riddle about their physical location in the form of a poem. The riddle, if solved, would have the player bend down and look through an opening under the arm of the statue toward the ceiling in such a way that a particular word and icon were in sight and could be entered as a short answer question, again allowing the system to determine their location. We called this mechanic *reinterpreting space* and applied it to other puzzles later in the game.

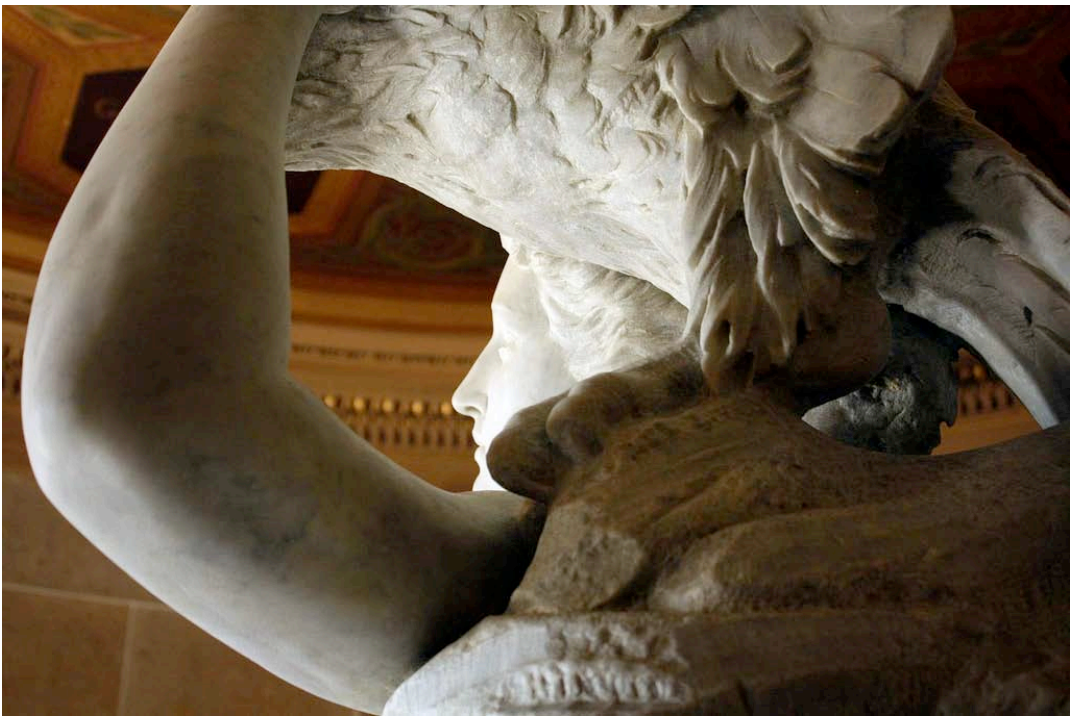

Looking through the arm of *The Genius of Wisconsin* to frame the art on the ceiling.
Image copyright: Ron Wiecki

In other segments of the game, students would *drag and drop onscreen components* to match their related physical positions demonstrating not only comprehension of the meaning of those objects but also verifying they were in the space. Later they would *apply a cypher* to the state constitution, decrypting a hidden message from the missing Dr. Hernandez. For the climactic scene, they would go outside the building and mark 3 points on the map to *triangulate* the doctor's current location, were we had placed a physical *geocache* to discover. As the design was nearing completion we began to notice that we were building an entire vocabulary of mobile game mechanics.

| Mechanics and verbs | Location detection |
|---|---|
| unknowing actors | branching choices |
| situated riddles | short answer input |
| reinterpreted original space or artifact (poems, cyphers, etc.) | drag-and-drop position matching |
| position triangulation | |
| geocache discovery | |

Table 1 - Design vocabulary after *Saving Dr. Hernandez*

The game was never tested with a live middle school aged class as we originally intended. It was simply too rough around the edges for general use. We did however play-test numerous times with ourselves and with Seann and Chris's children. As early prototypes should, *Saving Dr. Hernandez* generated a host of questions and next steps for not only for our game designs but also the technologies we would need to instantiate them. By those terms, it was easy to call the project a success, but it was a bit disappointing that we did not have a traditional design and control study on state standards completed as originally hoped. We wrote up a short post mortem paper about the project and moved on.

About this time, Kevin Harris joined the team and we decided to try a more experimental approach for our next design. It seemed that something classroom-ready was still a way off and we needed to give ourselves time to experiment and discover without committing to a large design, which rapidly slows as the content grows. Along these lines, we were curious about the design around alternate reality games such as *I Love Bees* (Microsoft & 42 Entertainment, 2004) and *Year Zero* (Nine Inch Nails & 42 Entertainment, 2007), deciding to make an alternate reality game that could be released to a public audience in episodes. That way we could have the advantages of a larger design (not have to reinvent the back-story for each game experiment), but stay with small, iterative releases that did not have complex interconnections. By releasing episodically, the team could also be split so one half was polishing an episode while the other was building the technology for the next. From an education point of view, alternate reality games seemed to point toward the kind of persistent, casual, informal learning environments we had always imagined.

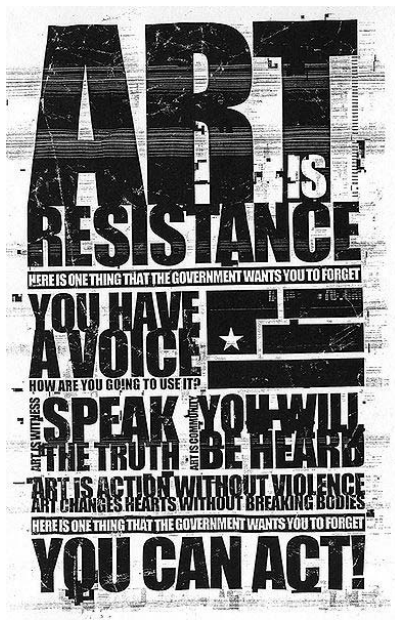
Poster for *Year Zero (Nine Inch Nails and 42 Entertainment, 2007)*

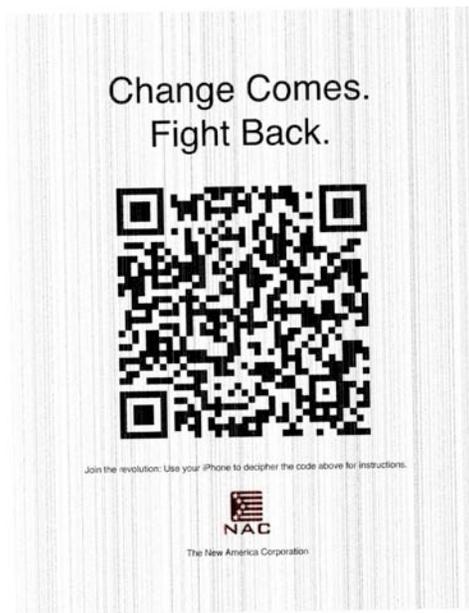
Poster for *New America Corporation*

Using ARIS 0.1, a port to HTML that would run on the iPhone, we built 3 episodes of a game called *The New America Corporation* or *NAC* for short. The game was designed to pull in members of the general population though cryptic posters placed around Madison that would direct them to contact the design team if they could determine it's meaning. We would respond with an iPhone app attached to an email, which would walk them through a psychological profiling process to determine their trustworthiness to the NAC and bridge them to episode 1. In the following episodes they would perform a set of tasks for the NAC that would slowly reveal the back-story of the game by having them perform various tasks around Madison.

The game functioned as a thin veneer of alternate reality that aimed to affect the player's normal lives. For example, tasks given by the NAC would take them to local eateries like Dotty Dumplings for a burger, or the Chazen Museum of Art on a recon mission that they could do while out with friends. In the final episode, players were to find a particular farmer at the Saturday farmers market, whom the design team had already briefed, and pick up a flyer that contained information about a competing faction that needed to be eliminated.

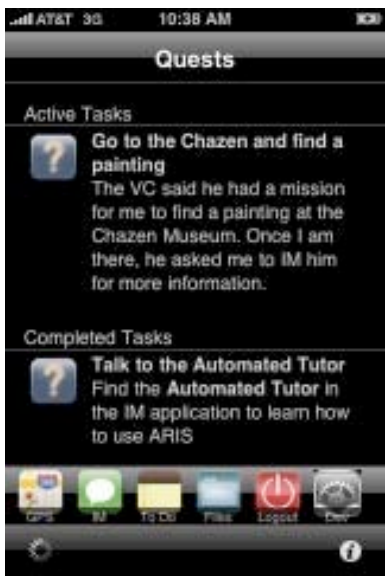
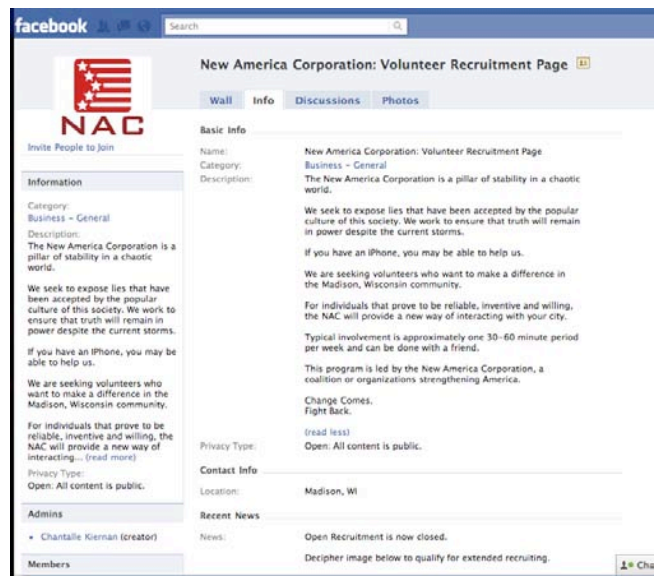

Screenshot of ARIS 0.2 running *NAC episode 1*    The NAC facebook page

The design team built a NAC facebook page and set up a few NAC Gmail accounts then ran a small pilot that involved 3 semi-suspecting friends. Later, a dozen or so students of an educational media design course, C&I 614, played through episode 0 and 1 and provided feedback. In general, there was a lot of interest in this kind of game but the team felt that the episodic process was still too slow. We hadn't really learned any new mechanics from the NAC project and our audience was unclear. "How many people would really even play a game like this?" we asked. The project was never implemented publicly and we transitioned again.

## Distributing Design

With an $8,000 seed grant from UW's Division of Information Technology - Academic Technology group, Kevin and I started work on what would be called ARIS 0.3. Peter Debbink joined the group and we set our goal: design and implement a tool that could be used by education students to produce simple mobile learning activities. We were able to build a new interface for the editor, which improved elements of the UI for placing locations and making links between the game objects. After documentation and design exercises had been created on a wiki, we facilitated a set of workshops in Dr. Squire's C&I 614 course for novice education game designers. The students began working on their games.

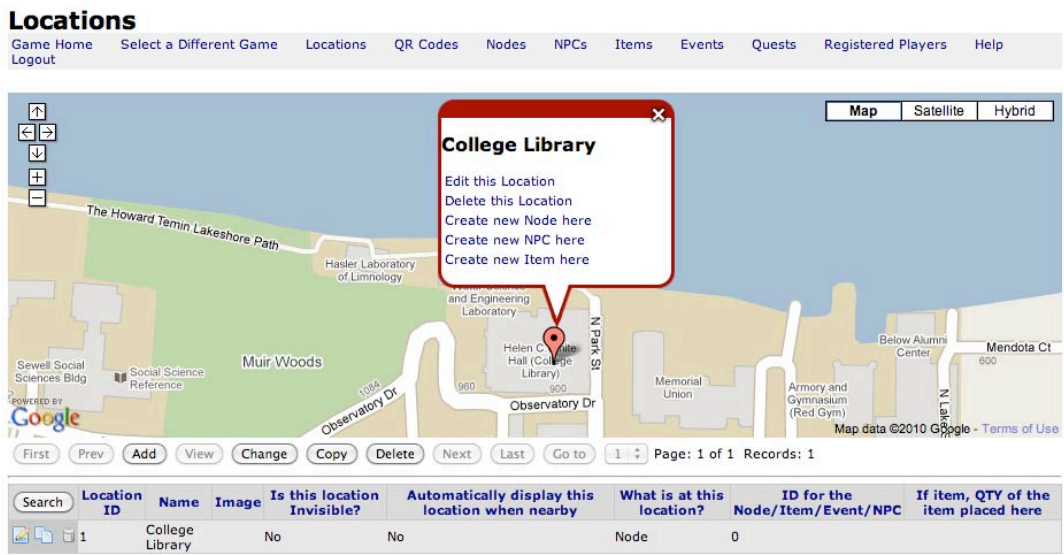

A screenshot from the ARIS 0.3 Editor

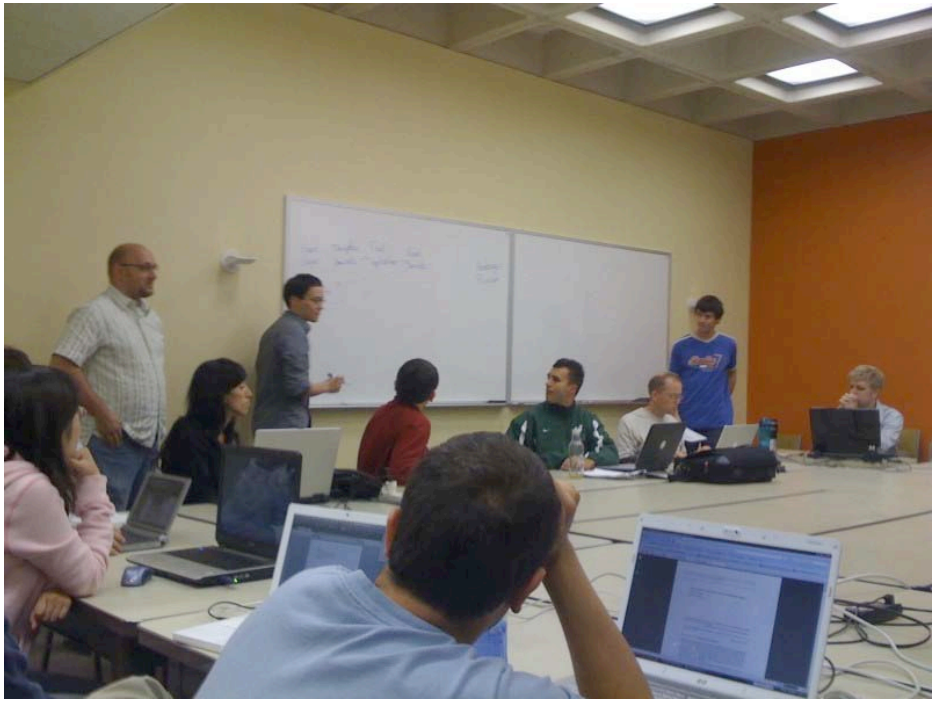

Kevin Harris, Seann Dikkers and Peter Debbink facilitating the first ARIS workshop for students in C&I 614

The team was excited by the idea that if many people were making these kinds of activities, we were likely to see some really interesting designs emerge and continue developing our understanding of mobile game mechanics. However, when we debriefed the semester long project, our conclusion was that game design is not something easily taught and the technical hurdles were not making things any better. Of the ideas pitched by the class, many of them were simple point-to-point walking tours and the more interesting ones were out of scope for what ARIS could do. The question that haunted us most was, "even if we build this tool for a general audience, will the resulting games be any fun to play?"

During the following phase, focus turned back into the team as the primary source of designers, though we folded in a few new faces. With additional funds provided by an ongoing MacArthur grant with Dr. Squire, we began another round of design exploration.

Matt Gaydos first suggested a game design jam, and the idea caught on. If a group would meet weekly to produce small prototype game designs, we should quickly learn what kinds of activities are interesting and should be incorporated into a general-use authoring tool. So GLS associates Jim Matthews, Ryan Martinez, John Martin and visiting scholar Muhammet Demirbilek joined the team as regulars and a few others such as Yoonsin Oh, Sarah Chu, Crystle Martin, Meagan Rothchild and Mark Wagler joined us a few times during the Spring 2009 Mobile Design Jam.

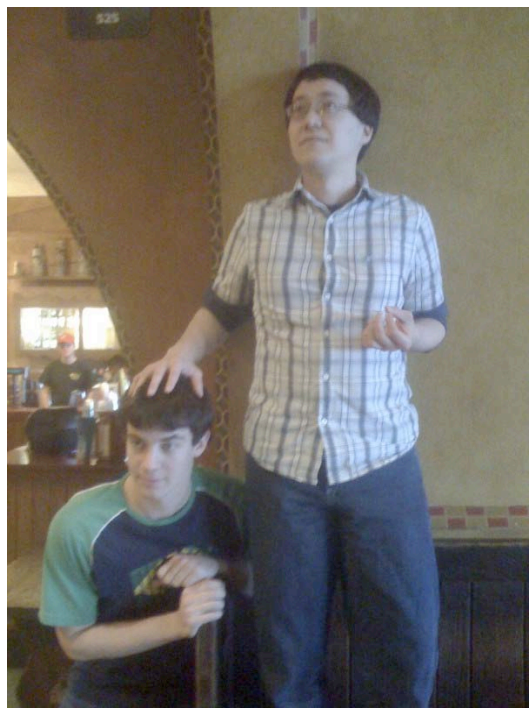

Example image captured during play testing *Reenactment* (Jim Mathews), a game designed during the Spring 2009 Mobile Game Jam

During 16 weeks, at least 21 mobile game designs[2] were implemented, play tested and evaluated using a variety of light-weight technologies. The general strategy was to meet weekly at the Memorial Union, cycling agendas between debriefing and play testing. With every 2 week iteration came a theme that served as a starting place for game ideas from individuals or small teams. The standing goal that ran through each design was to highlight a particular location-based mechanic and see what could be fun or interesting about its application. Most of the resulting games were short (<15 minutes of play) and involved the Memorial Union and its guests directly. Near the end of the semester, with summer fast approaching, we held a public vote to determine which ideas would be implemented in ARIS to be played at GLS 2009. We also achieved what we set out to do, growing our understanding of promising mobile mechanics and verbs through worked examples.

---

[2]See Appendix for a list of designs

| mechanics and verbs | location detection |
|---|---|
| unknowing actors | branching choices |
| situated riddles | short answer input |
| reinterpreted original space or artifact | drag-and-drop position matching |
| position triangulation | QR Codes |
| geocache discovery | GPS / WiFi |
| scanning QR code geocaches that link to virtual objects | |
| collecting photos, videos and sounds | |
| collecting/trading virtual items | |
| modifying a shared virtual world | |
| documenting a reenactment | |
| recruiting additional players from environment | |
| seeing/racing other players on map | |
| location avoidance | |

Table 2 - Updated design vocabulary after 2009 game jam

## Investing into Open Education

During the same time period that we were facilitating design workshops for the game design course, Chris Blakesley wrote a paper that was accepted to the Open Education 2008 International Conference, putting us right in the center of the OAR community and vision. During our demo and presentation, I announced that ARIS would be licensed under the MIT license[3], making it free for all use, modification and even commercial derivatives with no reservation or even attribution. The 20 or so participants applauded the generous decision and I knew it was the right move if ARIS would ever succeed.

---

[3]See http://www.opensource.org/licenses/mit-license.php for details

During the lunch that followed we received a few tips from participants who had more experience with launching open source projects. They told us some best practices about making ARIS available for outside contributors and few people at the lunch table that said they were interested in participating. Within a few weeks, http://code.google.com/p/arisgames/ was setup with every line of code written for every version thus far for anyone to download and start tinkering with. We also soft-launched http://arisgames.org/editor for designers to start playing and posted the documentation written for the game design course to support them. Even though the team had established upfront that anything created during the project would be given away and that grant funding would likely never result in a sustainable tool, the trip to Utah for the Open Education conference and posting the code to a public repository showed we meant it. There was no turning back. ARIS was on the Internet.

## Releasing Early and Often

In New Mexico, GLS alumni Chris Holden, with colleague Julie Sykes, began working on Mentira, a game to teach intermediate Spanish concepts. He used only the semi-built version that was posted to the Internet after the conference, bugs and all. In total, the game and surrounding activities take place over the period of 3-4 weeks, replacing one oral presentation requirement in the Spanish 202 curriculum.  Possibly due to a background in mathematics and previous exposure to a mobile design platform, Chris was able to easily navigate the complex data structures of the ARIS authoring tool well enough to engage in a production ready game, even requesting access to the raw SQL interface for final details. Their implementation was dozens of times larger and more intricate that anything designed in Madison at that time. This description can be found at their web site, http://mentira.org:

> Mentira is a place-based augmented reality game using the Augmented Reality for Interactive Storytelling (ARIS) platform, developed at the University of Wisconsin-Madison, for use on ipod touches and iphones.
>
> Our game consists of somewhere near 70 pages of dialogue and informative text, almost all Spanish, about 150 photos or pieces of still visual art, and four short movies. It is somewhere near 20 times larger than any other game or tour developed for the ARIS platform.
>
> The game is a murder mystery, consisting of current and prohibition era fictional events, all set in the Los Griegos neighborhood in Albuquerque. The basic structure of the game is directed conversations between the player and fictional characters (Non Playing Characters or NPCs) concerning the murder and its solution. Each conversation is situated at a particular place and time in the game's narrative, somewhere between reality and fiction.
>
> The basic dialogue game mechanic of the game will feel somewhat familiar if you've ever played any of the Carmen SanDiego games, although ours does not hinge on historical trivia or logic puzzles.

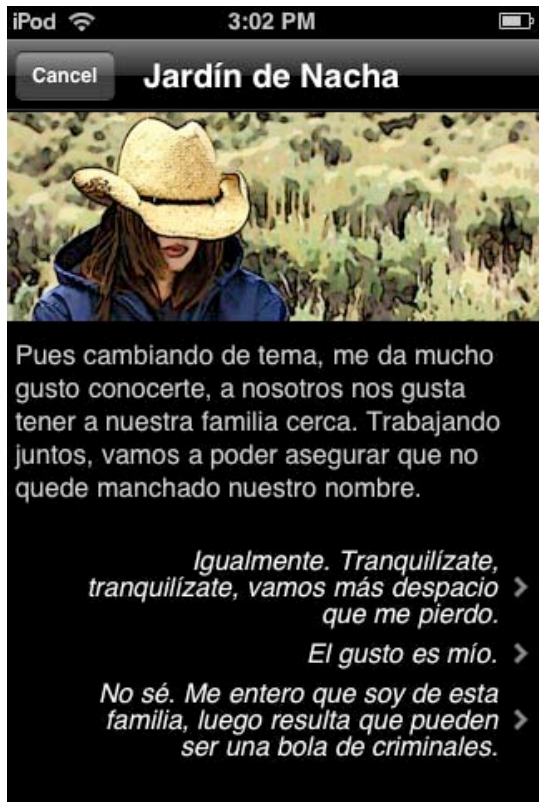 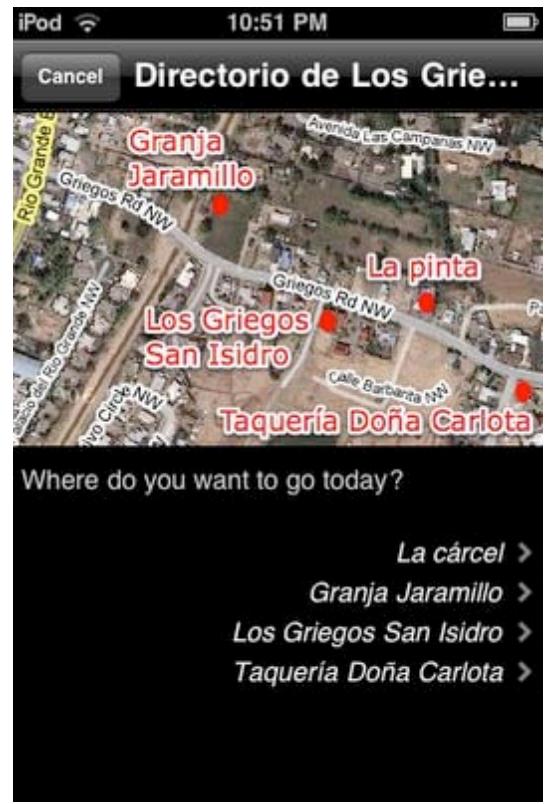

Screenshots from Mentira, a Spanish language game written in ARIS 0.4

This was the first example of how releasing early and often is a strategic move in open software production. Not only had the group in New Mexico provided our flagship demo, but they had learned implementation details along the way that our Madison team would have never discovered. Sadly, the ARIS team had no concept of what had been achieved in NM until the groups finally met up at GLS 2009.

In what is becoming a habit, the ARIS team set high expectations for what would be shown at the yearly GLS conference. In 2009, with the game jam designs being polished and the 0.4 ARIS release almost completed, we decided to facilitate a session that would contain a live demo of one of our games. *Ghost Hunters* was selected as the best fit because it could be played indoors in the shortest amount of time. Adding two more members to the team, Thomas Henage, a physics PhD student and Brian Deith, a learning consultant for the college of Letters and Science, we polished and tested numerous iterations of the game, right up to the day before, in preparation for our first real public demo.

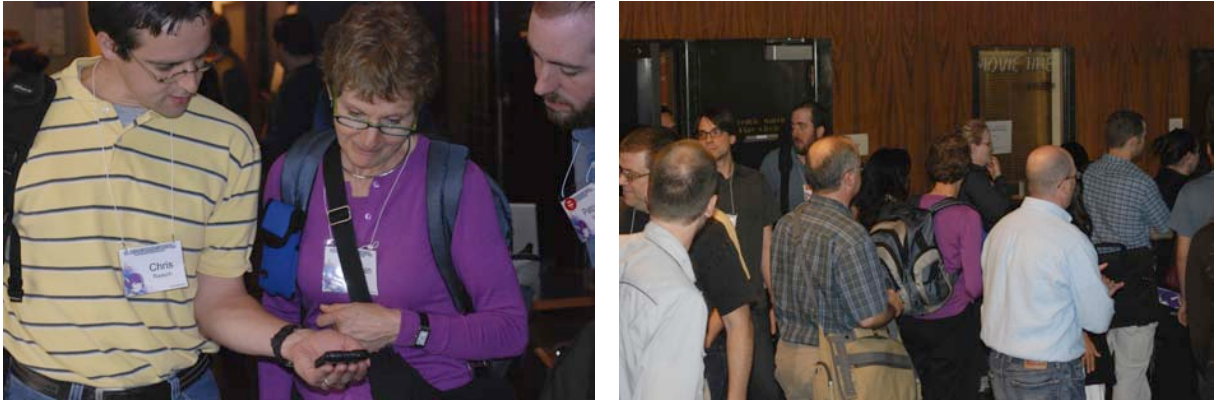
Participants using ARIS during GLS 2009

Though we had a few other designs that were played by a few people, the ARIS session featuring *Ghost Hunters* was a huge success. Near 30 participants created a wonderful ruckus as they left the presentation hall and started wandering around the building, laughing and talking about the virtual trinkets and voodoo dolls they were discovering by scanning QR Codes hidden around the Memorial Union. Each team played through with only minor technical issues. Because iPhones were becoming more popular, most were able to play with their own personal devices, but we provided five loaner devices to fill in. There was enough time to come back together and debrief at the end.  The main message we heard is that the activity was interesting and they wanted more. We had met our mark.

## Publicly Launching ARIS and A Community Around Mobile Learning

Following GLS 2009, Thomas Henage began work as a paid developer in addition to myself, Kevin Harris and Brian Deith. Additional funds from Dr. Squire enabled another round of design and development, this time returning to the goal we started with: A stable game engine with an easy to use authoring tool. During the following months we sketched dozens of designs, made mock-up UIs that were carefully user-tested and even recorded a few screen movies to distribute to our growing online community for feedback. Having a student dedicated to the authoring tool programming increased our development speed dramatically. During this time we also began work on the 1.0 version of ARIS that would be released to the Apple App Store.

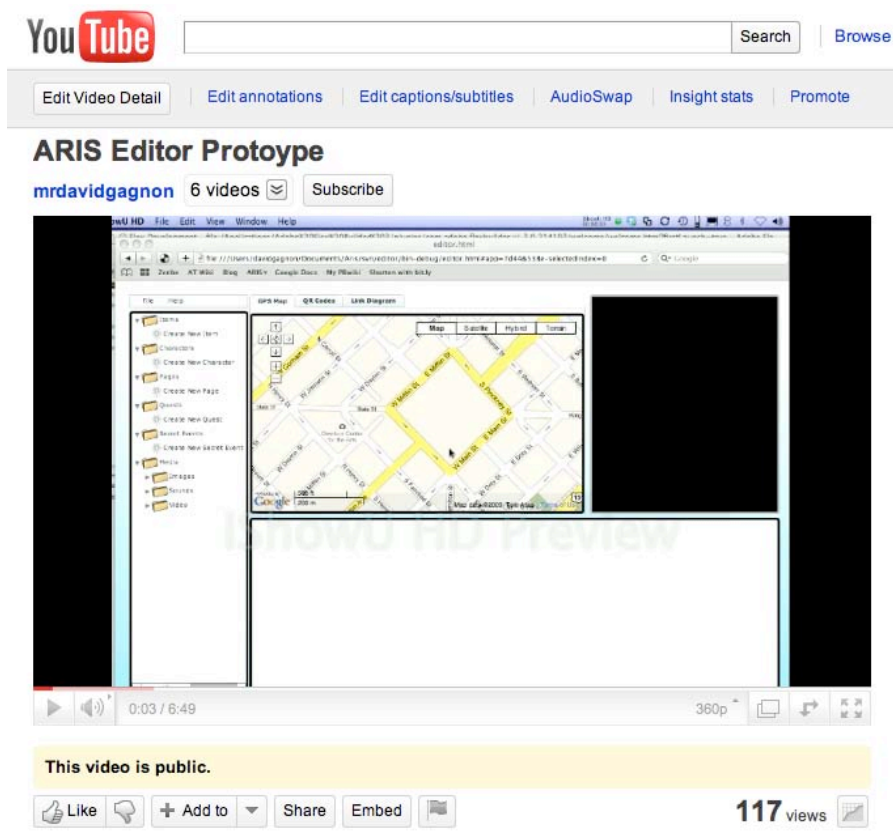

Posting authoring tool UI mockups to YouTube to collect user feedback

During the following season we had a number of severe setbacks. First, Thomas decided to leave the PhD program and would need to be brought on as a Limited Term Employment if he were to stay in Madison and work with ARIS. Despite the fact that we had the funds needed to support him for a year at double the rate he was currently being paid as a student, the political environment at DoIT was not supportive of the idea and we had to let him go. We transitioned the authoring tool development to Learning Solutions, a professional software development group at DoIT that I was used to working with for other game projects. Sadly, numerous problems arose immediately. The programmers assigned were confused by the team's design methods and ad-hoc membership. They were not comfortable hosting code on public Google repositories or using the team's 3rd party project management tools. And though the final builds from Thomas were near complete, the new developers were not comfortable starting with a student's code and convinced us to start over. It was a costly set of negotiations and the hourly rate for development had more than quadrupled by the change to senior programmers.

During this turmoil, Kevin Harris, a key developer and the only other team member that understood the entire ARIS architecture in entirety, pursued another opportunity and left the team. Around the same time, however, John Martin was hired on full time and Chris Holden began meeting with us weekly via iChat. The team composition had changed dramatically and we were now working with outside contractors for development.

As development crawled along, attention turned to building community around mobile learning designers. Despite excitement in publications like the PEW Internet mobile survey (Lenhart, 2009) and the New Media Consortium's Horizon Report (Johnson, Levine, Smith, & Stone, 2010) about mobile learning, I concluded that no conference or organization in the US had really become a home for people interested in the topic. So starting locally, we announced that an event was going to be held on campus for all interested instructors, designers, IT support staff and instructional staff. With the announcement came a flood of supporting emails from all over campus and from all levels of my management. Nearly 50 people from UW Madison were in attendance and shared projects and concerns from their varied perspectives. This single move had caught the attention of the entire campus and placed our team in a central spot to hear everyone's new ideas.

Scaling up, we began talking about doing something on the national scale at GLS 2010, creating a gathering for people interested in mobile learning to meet each other, and share ideas, projects and inspirations. After pitching the idea with Dr. Squire and then Dr. Constance Steinkuehler, the conference chair, our team began planning and meeting via Skype with Eric Klopher's team at MIT and Coleen Macklin at Parsons school of design. We wanted to create an event where every participant was a designer and all ideas were given an ear. This is the description that was posted to the GLS website:

> The Mobile Learning Summit is a pre-GLS, 3 hour time of brainstorming and networking, hosted by Games Learning and Society and UW Madison's Academic Technology Department.
> 
> This event will be unlike most conferences you have attended. From the onset we have designed our time together to meet these two goals:
> 
> 1. Encourage authentic collaboration and community around real people in the domain of mobile learning
> 
> 2. Create an active, risky environment where every participant is an expert and emergent ideas are expected

After talks with Kurt Squire here at UW Madison, Eric Klopher's team at MIT and Coleen Macklin at Parsons we have decided to do something fun, bold and messy: We are going to make the GLS-MLS a facilitated mobile design jam.

If you haven't been to a design jam before, the goal is to produce a tangible artifact in a short amount of time. Along the way we will be working together in teams and hopefully getting to know each other and some of the people that have gone before us. All of the designs made during this time will be under the [Creative Commons Attribution License](#) for anyone to pick up and move forward.

**Registration is Free** and done when you register for the GLS. Coffee and breakfast will not be provided.

If you want to start getting to know each other right now and share what you are thinking about, join the [GLS-Mobile Facebook group](#).

Risky as it was, Coffee and Breakfast were provided care of Sean Michael Dargan and over 120 people attended the event. Coleen did most of the facilitation and Dr. Squire and I provided some context to the event and mobile design history. During the 3 hours, teams of 6-8 each generated hundreds of game ideas on post it notes and escalated one or two of those ideas into a full storyboard. Seeing high school teachers collaborating with scholars like Henry Jenkins and Eric Klopher was rewarding to watch.

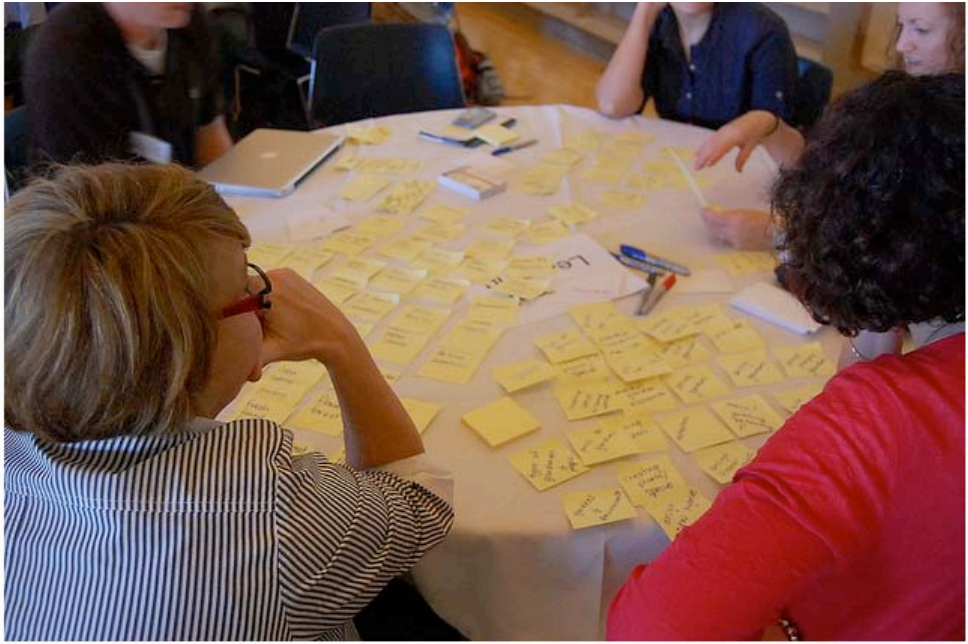

A photo from the GLS Mobile Learning Summit with comments from flickr. Each post-it note represents one game idea being pitched to the group.

While we were planning the Mobile Learning Summit, Chris Holden and Jim Matthews successfully ported a previously created situated documentary, *Dow Day*, from archives of the MIT Outdoor AR engine into ARIS. Ron Cramer, a colleague at DoIT constructed a historic UW *Madison campus tour* in ARIS based on a script he wrote and edited using images from the State Historical Society and University Archives. Our team also quickly ported the metal collection and smelting game, *Steel*, from the 2009 game jam into the new software. So with 3 games ready, ARIS 1.0 was released to the App Store and publicly launched at GLS 2010 with hundreds of fliers and numerous mentions from keynote speakers during the event. Over 200 copies of ARIS were downloaded in less than a week.

The final leg of the GLS conference was just as important as the first, even though the crowd was a tenth of the size. We were invited to host a session at the "educator symposium," a daylong mini-conference following the main GLS targeted at K-12 teachers. During a two-hour workshop, we had participants use the brand new, flash based authoring tool finished by LS only days before. They created two simple mobile activities and went in the field to test them. Surprisingly, everything worked perfectly. As we expected, the participants requested ongoing access to the tool. So with numerous warnings about the software only being in alpha, we took down all their email addresses and added them to a private Google group where information about authoring ARIS games and discussions was hosted. We also invited them to tell their friends to apply to the alpha[4] testing program as well. ARIS was finally at 1.0 even if the authoring tool was still in alpha. Once again, we had a lot of attention, but no remaining budget.

## Distributing Ownership, Investment and Commitment

Small (or non-existent) budgets can be a refining process for experimental ideas like ARIS. It was during this time that we had to wait and see if people would start using it and what they would use it for. This time reminded the team that if a project like this was going to survive it either needed to generate revenue directly, which was impossible because we had already given it away for free, or it would need to create partnerships with funded projects that could sustain its continued development. There were things we could do to encourage the latter.

---

[4]Alpha is a term to refer to software that is not yet feature complete and often buggy. Beta is the state in which all the features are implemented and testing begins to improve stability.

The first opportunity for a partnership came when Sabine Gruffat and Bill Brown, professors of Communication Arts at UW Madison, asked if I could help them with a locative media art project they were doing in NY called Bike Box. They had a very small budget and my position at DoIT allowed me to work with them for free, adapting the ARIS code to the specific project's requirements. It was exciting to see how easily and rapidly ARIS could be re-purposed for something as specific their concept, Bike Box. Within a half-day of work we had a working version of an app that looked nothing like ARIS and met all the goals of the artists. I was able to attend the opening of Bike Box in Brooklyn and met numerous artists who were also curious about locative media but did not have the technical ability to write their own mobile applications. From that show alone two other collaborations have taken place and we eventually did a similar process to create a custom version of ARIS for Mentira.

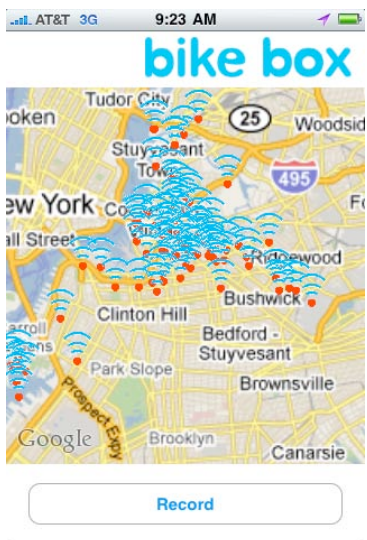
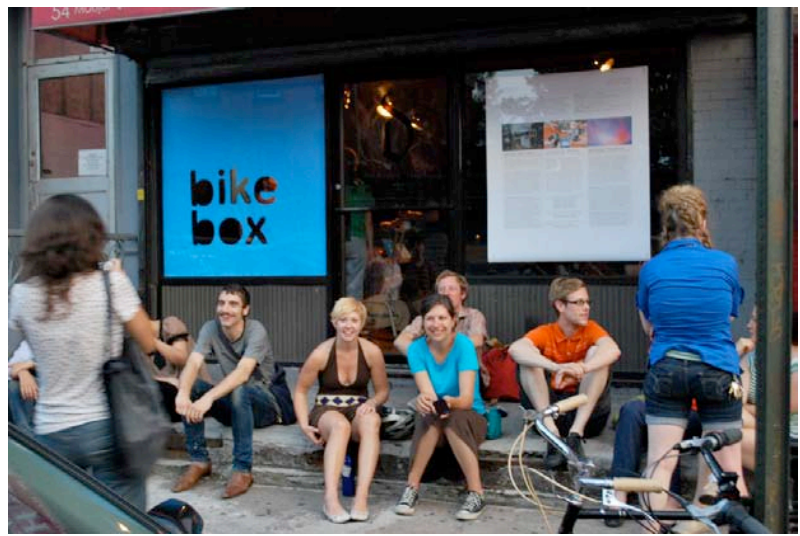

A screen shot from Bike Box, a branch of ARIS

Hipsters outside the Devotion Gallery where the Bike Box show was launched

Unfortunately, unfunded projects like were not likely to keep ARIS moving forward forever. We needed to learn how to attract contributors, collaborators and financial donors. Even open-source software is produced in majority by paid developers (Fogel 2010). ARIS needed to posture itself to be picked up by a diverse group of interested parties that will contribute back to the out of generosity and gratefulness that 95% of what they wanted to do has already been developed for free.

To this end, a number of inexpensive efforts began to collect attention and lower the barrier for contribution. For content creators, the arisgames.org website was overhauled visually using a $25 wordpress skin and better training videos were produced. We set up a Google moderator site to collect ideas for new features and report bugs and opened up the private Google group for public use.

For developers, we rebuilt the code repository with screen shots and dozens of pages of technical documentation. All of the UI sketches and mockups were moved from private wikis to this public space and our issue trackers, which show the ordering of tasks in the development queue, were given public access so anyone could see exactly where the project was headed and what bugs we were aware of. A "How Do I get Involved" document outlines points of entry at many skill levels while explaining the current governance structure for the project. In general, the goal was to be as upfront and transparent as possible about the project's roadmap, status and governance.

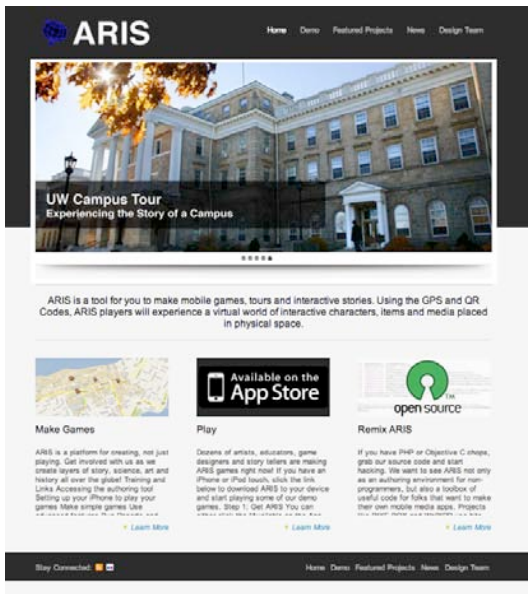
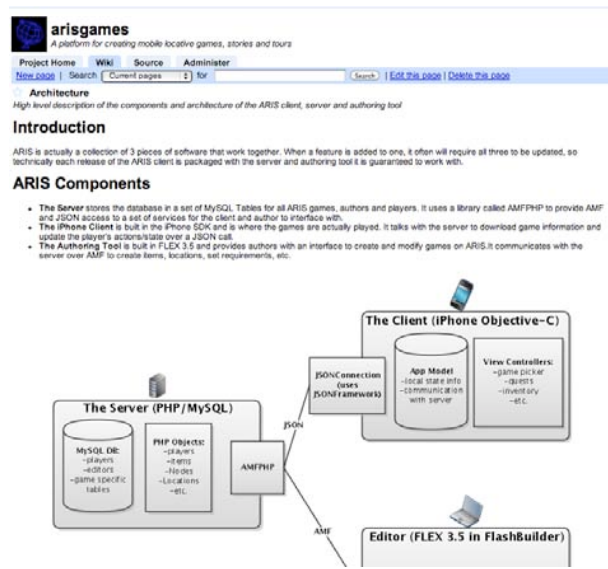

arisgames.org home page showing featured projects and ways to participate

Technical Documentation on public code repository

While not directly in response to the efforts outlined above, a number of collaborators have surfaced in proximity to the changes. Dee Johnson, a high school senior at Shabazz HS, formally joined the design team as an elective in his program and is working on the editor software while learning Flex. As of December 5th, he has made 14 commits to the public code, making visual improvements and fixing bugs in the authoring tool. High School computer science teachers at both West as well as LaFallettte are currently creating programs so their students will contribute development work to ARIS as part of their curriculum. Also, a team of instructional technologists at Amherst College has begun work to allow the current version of ARIS to function untethered from the Internet for single player games.

What may have been the most strategic use of a week since GLS 09, in early November 2010, the whole team blocked off three full days to get together in person and do another game jam. Chris Holden joined us from NM. Daniel Gosch, an undergrad student who was using ARIS in a course in CO, came as well. We reserved a cramped room at the Memorial Union and worked for three days straight, projecting new design ideas on the walls and covering the whiteboards with post-it notes. In these few short days, using a modified version of the Scrum development process with only *2 hour* iterations, the team produced 5 new game prototypes[5], improved the existing Dow Day game greatly and made dozens of changes to the authoring tool that were released to the public. Here is how each day was facilitated:

1. The day was divided into 2 hour-long iterations, from 10am-12pm for example.
2. A whiteboard had the categories 'current', 'backlog' and 'icebox' written on it. On the sides were 'done' and 'dead' areas.
3. Each member of the team wrote a single task they would like to perform on a post-it note and "pitched" the idea to the other members.
4. Short-term, small teams would form around ideas that sounded the most compelling and feasible and placed those tasks in the 'current' or 'backlog' category on the whiteboard.
5. Ideas that generated less excitement slipped into the 'icebox' or 'dead' idea piles.
6. Each team would perform focused work for the length of the iteration.
7. When the time limit was up, the team debriefed, adjusted goals based on past velocity and picked the next set of current tasks. Only 'delivered value' kinds of things, not research, chores, etc. counted toward progress.

In addition to the design work, we hosted two public events, an ARIS workshop and a play test of the games made that week[6] in order to involve local faculty and give us a specific deadline.

---

[5] See the appendix for a complete list of games produced during the 2010 game jam
[6] More details are available at http://arisgames.org/2010/11/madison-design-jam-postmortem/

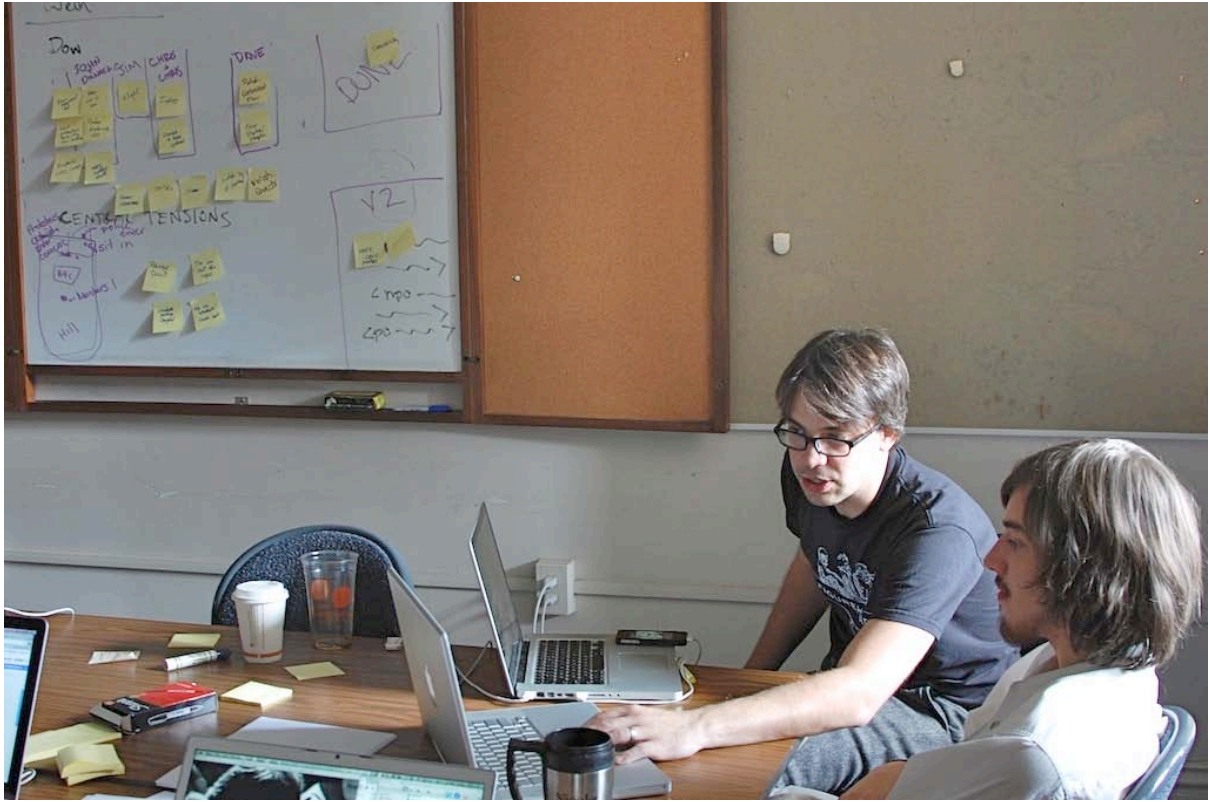
Working alongside Dee during the November 2010 Game Jam

# Current Status of the Project

### ARIS Software[7]

The authoring tool provides a web based, drag-and-drop interface for authoring virtual items, virtual characters and locative media. It is developed using the Adobe FLEX 3.5 SDK, uses a number of supplemental open source libraries and the Google Maps API. Communication with the server is done through the AMF3 protocol. Though It has a number of bug fixes and game administration features to be completed before moving into beta, it is useful as is.

---

[7] For the most current information and demos of the ARIS software, visit http://arisgames.org

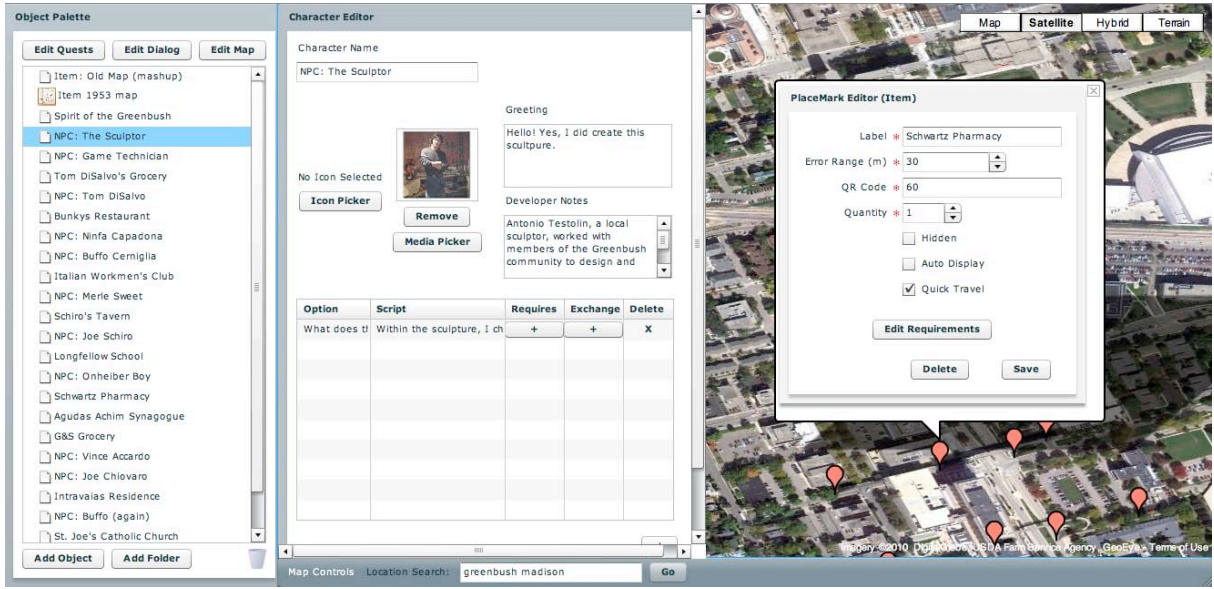

A screen shot of the publicly available authoring tool as December 5th, 2010

The iPhone Client is now at version 1.2 and has become increasingly stable. It is written in Objective-C and communicates to the server using the JSON protocol. Additional open source libraries provide QR-Code identification, robust network infrastructure and JSON parsing.

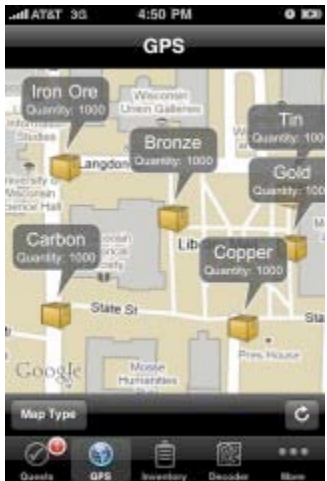 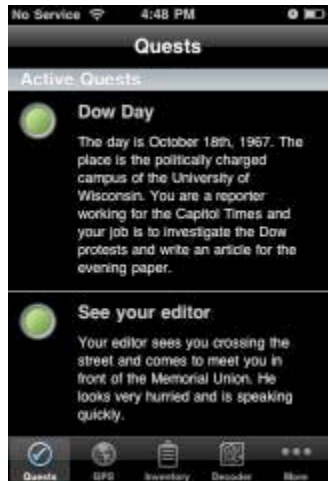 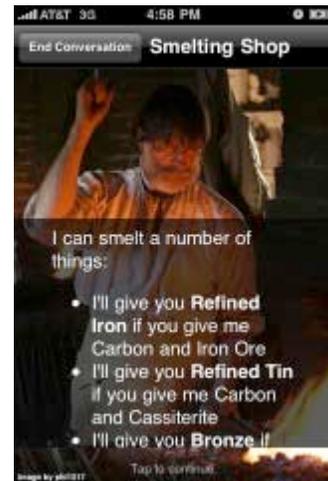

The map interface — Quests — Conversations with NPCs

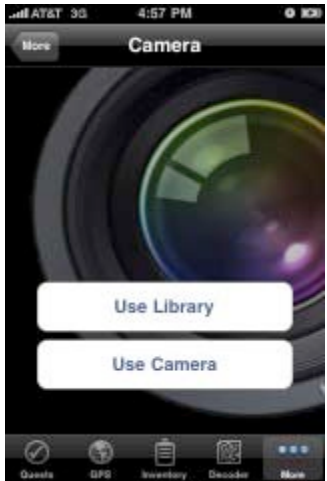 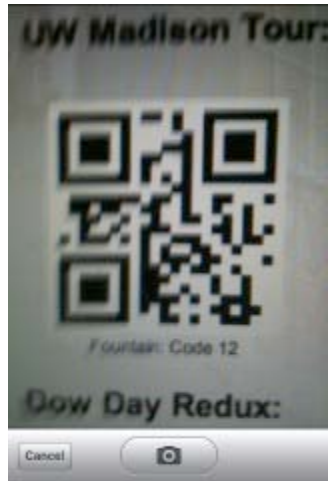 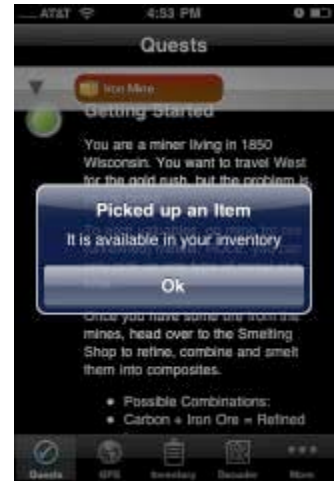

Camera            Scanning QR Codes            Inventory

## Web Site Use

Visits to the http://arisgames.org website are tracked by both the *Site Stats* module for *wordpress* as well as *Google Analytics*. Since we only began tracking usage with Google in August, missing the key launch event at GLS 2010, historical data will be used from Site Stats, but blended with more in-depth usage data from Google.

In general, ARIS is experiencing rapid growth both nationally and internationally since June of 2010. Website page views have increased an average of 34% per month over the last 6 months, peaking at 4,384 in November, an average of 146 per day. This equates to 1,704 complete visits to the site by 1,041 unique users per month. Visitors spend an average of 3.5 minutes on the site with the most popular pages being the home page and demo video.

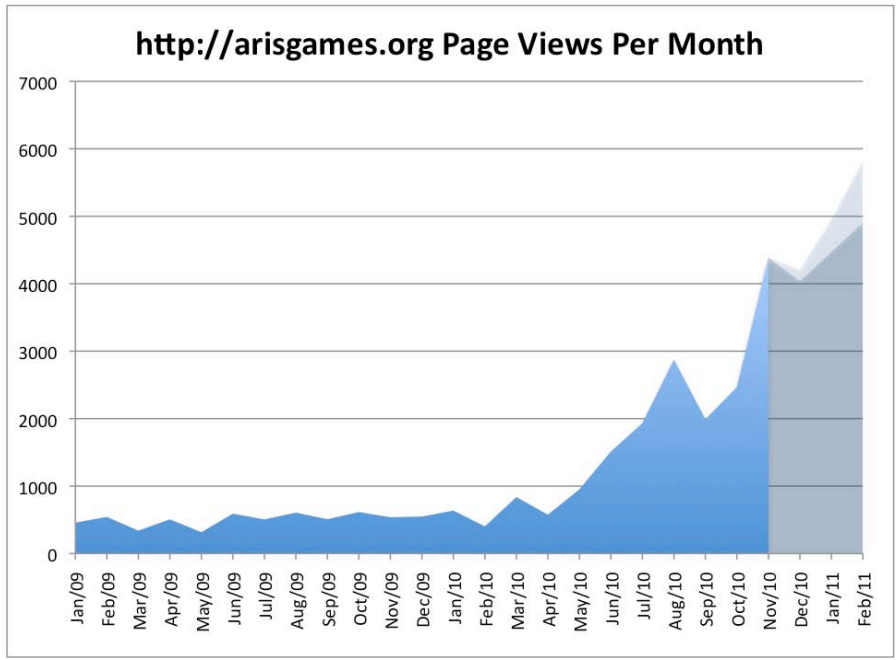

Page views for the arisgames.org website

numbers past November 2010 are predicted using growth and linear models

While the United States is by far the largest source of traffic to the site with 1,193 complete visits during November, 131 came from Australia, followed by Canada (46), Spain (33), United Kingdom (32) and the Netherlands (29).

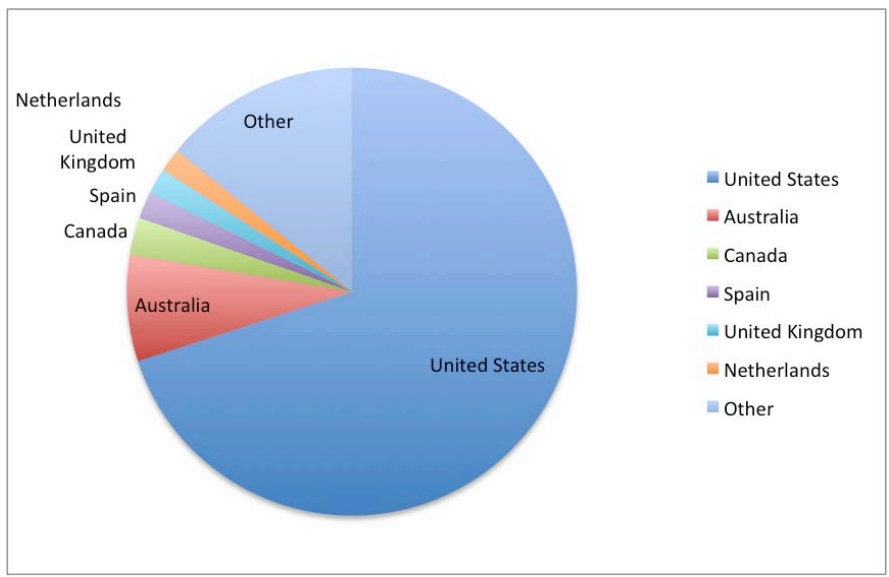

## Authors, Players and Games

As of December 5th, 143 unique authoring accounts, 659 public player accounts and 221 games have been created. Of those games, 31 contain more than 25 game objects and 43 have been opened by more than 5 unique users. Most of the games appear to be small tests.

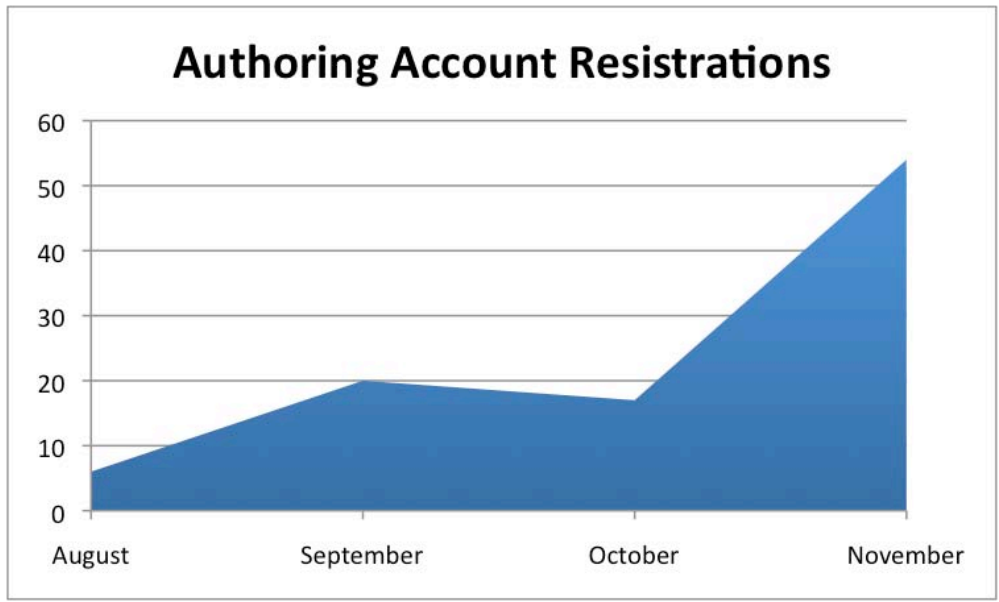

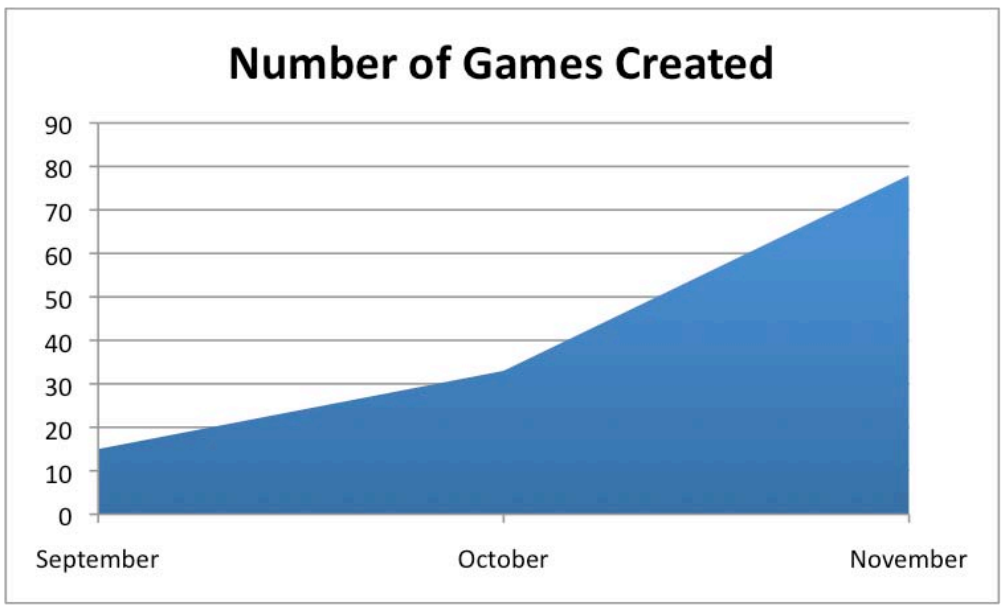

## Papers, Press and Presentations

ARIS has generated a fair amount of press along with numerous writing and speaking invitations. It was featured in the 2010 Horizon Report (Johnson, Levine, Smith & Stone, 2010), The Chronicle of Higher Education (Li 2010) and Spotlight on Digital Media and Learning (Jackson, 2010). To date, two papers from the ARIS team have been published about its use, one for Educause Quarterly (Gagnon, 2010) and one for the International Journal for Game Based Learning (Holden and Skyes, In Press). In addition to dozens of conference sessions performed by all members of the team, I have personally received numerous speaking invitations because of ARIS work.

## Collaborations

As of December 5, 2010, ARIS has been written into at least 7 grants with collaborators at Georgia Institute of Technology, The University of Wisconsin Arboretum, Rochester Institute of Technology, the University of Buffalo, the Mobile Learning Institute, the Board of Jewish Education of Greater New York and Old Dominion University.

Members of the ARIS team were recently awarded $24,000 from the Mobile Learning Institute for the development of tools and curriculum to be used at the Smithsonian in DC and Field Museum in Chicago. There appears to be a recent trend for the ARIS team to help create workshop programming with informal learning institutions. Current negotiations include work with the Library of Congress in DC and New Youth City in NY. Numerous groups are also currently using ARIS for facilitating mobile media design workshops without our paid support, including partners in Spain and France.

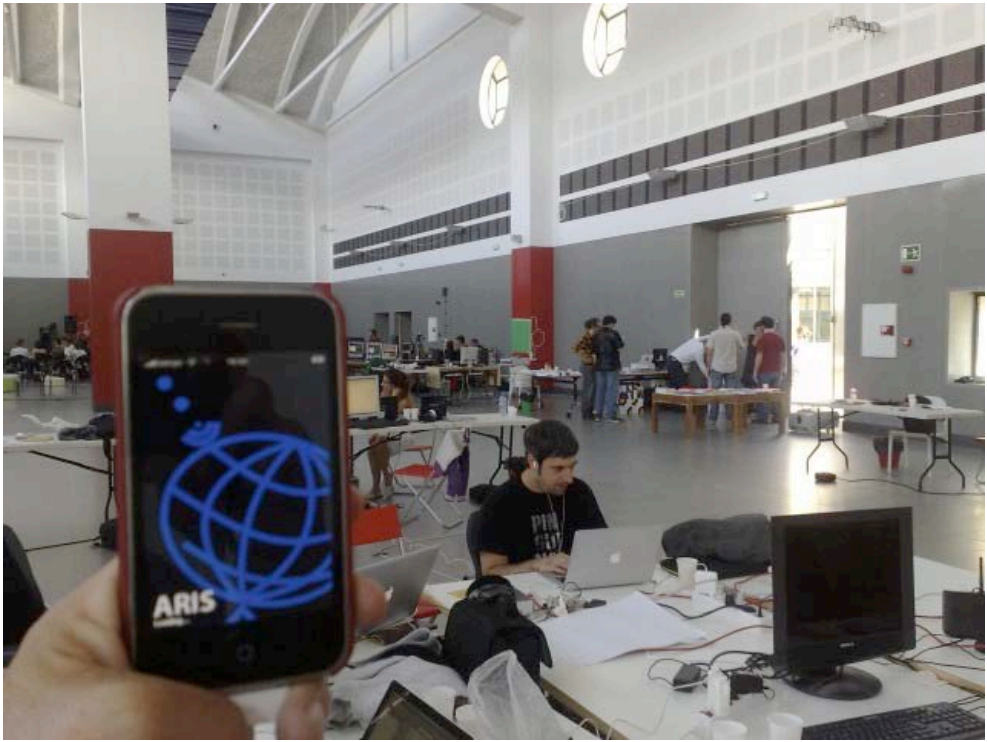

ARIS being used during a workshop at LABoral Art and Industrial Creation Centre, Spain

A second trend of projects involves ARIS as a mobile field data collection tool. Though only in prototype and awaiting funding, ARIS is being extended to provide robust social-casual gaming structures around collecting scientific data like bird sightings, water pollution measurements or geotagged plant identifications.

# Conclusions

ARIS is the union of game-based education with mobile media technologies and open access values. Its development has been iterative, conducted by a team of Games, Learning and Society affiliates in response to lessons learned by a community of global partners as well as a number of internal design jams and demos.

In addition to a software tool and user community, ARIS also has produced a number of curriculum artifacts such as Mentira that are still available for continued use and evaluation. Through rapid prototyping, ARIS game jams have contributed a formative set of mobile media game design verbs and mechanics that will inform future work. Finally, the ARIS team is beginning to do meta-design with informal learning institutions, teaching teachers how to facilitate student studio activities.

Looking at the increasing velocity of the last three years and the host of possible projects going forward, it is safe to say that this work has only just begin.

# Appendix A - 2009 Game Jam Projects

**Dinopet** (Chris Blakesley) - Find essential food for your Glyptodon or risk losing it. An experiment with obtaining a persistent virtual character and sustaining it with place-based resources.

**Unite** (Matt Gaydos) - In unite, two teams raced to take the correct photos and return first to the starting place.

**Library Fisticuffs** (Ryan Martinez) - Fisticuffs launched two teams of players from different points to solve location-sensitive riddles and return to the starting place first to win.

**pwned** (David Gagnon) - This was an experiment with an "interactive virtual object" masked as a mysterious wifi transmitter. If the player accepts the connection, their device is hacked.

**SMS Helix Riddle** (David Gagnon) - This activity begins with a riddle found in a SMS message that was intercepted by the players organization. The riddle leads to a particular building on campus and is answered by typing the correct address into ARIS.

**Indoctrination** (Kevin Harris) - An attack is imminent and the player needs to interpret the color of a beacon to understand the enemy communication.

**Saving Philip** (Yoonsin Oh) - A famous statue needs to be revived! Find the statue and correctly answer the questions to save him.

**Uncle Andrew's Inheritance** (Kevin Harris) stretched comfort zones, as players challenged strangers (with the names needed to win) to paper-rock-scissors competitions.

**Hobo Wedding** (author wishes to remain anonymous) challenged ethics and ingenuity, with the goal to "steal" items a bride would need - something old, something new, something borrowed, and something blue.

**Steel** (John Martin & David Gagnon) utilized ARIS as players competed to collect virtual resources and objects surrounding the Memorial Union. Players strategized to build strength for a culminating classic dice-based battle.

**Time travel puzzle** (Matt Gaydos) riffed on UW-Madison's alma mater in a time travel guessing game.

**Ghost Hunters** (David Gagnon) sent teams of ghost hunters (the players) to capture a troublesome apparition. Uses a story embedded in QR codes placed around the memorial union.

**Walls of the Union Riddle** (Matt Gaydos)

**Reenactment** (Jim Mathews) photos of different murals at the union must be reenacted by the opposing team and captured with a photo

**Rock, Paper, Scissors with the Camera** (Kevin Harris) - Groups of one, two or three strangers must be photographed doing actions form a list, then the photos are used to 'battle' the other team

**Weapons of the Union** (Unknown Designer) - Weapons are photographed in the union and brought back to a battle

**Food Groups** (Chris Blakesley) - Photograph food items around the union to create the meal that will build the best stamina and defense

**Judged Shot List** (Peter Debbink)- Photograph as many of the items on a list of 20 to score points in a time challenge

**Steel II** (David Gagnon) - Collect different ores from around campus and take them to the smelting shop near the memorial union to combine into composites of greater value

**Land Mines!** (David Gagnon and Kevin Harris) - One member of your team needs to make it through the rice patties of Cambodia without being destroyed by the hidden mines.

**Artifacts of a Tale** (David Gagnon and Chris Blakesley) - Items placed around campus tell a tale if you can put their meanings together. Players then move the items to give the story a whole new spin to the next set of participants.

# Appendix B - 2010 Game Jam Projects

## Games

**Contested Spaces (Chris Holden)** – A new lens to view your world. As you walk around campus, notice places where different interests come into conflict and comment on them by geo-tagged and shared videos, sounds and photos.

**Robots!** (John Martin) – A race of robots has been stranded on earth and you need to help them find their way home by taking photos of the robot icons that have been painted on roads and buildings around downtown Madison.

**Madison Tower Defense** (Daniel Gosch) – Unit 6, the package has been dropped at the locations marked by your GPS. Retrieve it and make your way to the extraction point before being spotted by the guards or sentry guns.

**Dow Media Tour** (Chris Blakesley and David Gagnon) – A dozen locations on campus have footage from the protests in 1967. Visit the places where photos and videos were taken and look out on the world from that perspective.

**Lakeshore Tour** (David Gagnon) – If only the trees could talk! From 100 years ago to current project, the walking path from the Memorial Union to the Hospital has so many stories to tell.

We remade a previously designed phone based tour using ARIS while beefing up the design in a number of ways.

**Dow Day Improvements** (All) - Added a new intro screen and cleaned up media and dialog.

## Software Improvements

**Editor Improvements** (Dee Johnson and David Gagnon) - Added the ability to author quests, conversations and use quantities in requirements. Fixed numerous bugs.

**Client Improvements** (David Gagnon) - Added quick travel, a method to virtually move to a location without having to travel physically and allowed the inventory to track items with quantities with quantities greater than one, effectively creating an attributes system.

# Appendix C - ARIS Version Descriptions

## ARIS 0.09

Platform: Flash running on desktop computer or MS Pocket PC portable device

Architecture: An XML file is read on launch. No connection to the Internet is required after that.

End of Line: Flash is not supported on the iPhone. Interfacing with camera and GPS near impossible without complex configurations. Pocket PC had a dwindling market share.

## ARIS 0.1

**Platform**: HTML and JavaScript generated by PHP and MySQL.

**Architecture**: Pages are generated at run-time from data stored in MySQL database on Internet server, requiring a constant connection. Connection to GPS enabled by providing variables through JavaScript from custom web browser on iPhone.

**End of Line:** A more robust PHP platform was needed to support multiple clients.

## ARIS 0.2

**Platform**: HTML and JavaScript generated by PHP and MySQL, using Framework engine

**Architecture**: Pages are generated at run-time from data stored in MySQL database on Internet server, requiring a constant connection. Connection to the GPS was done by accessing custom web browser variables through JavaScript.

**End of Line**: HTML and JavaScript were increasingly difficult to implement for our aesthetic. No camera access.

## ARIS 0.3.x

**Platform**: HTML/JavaScript and JSON generated by PHP and MySQL, using Framework engine

**Architecture**: For certain functions, HTML pages are generated at run-time. For client-side functions, a JSON dataset is generated and rendered on the client. This allowed us to have full connection to the iPhone UI and easily send client location information back to the server. In 0.3.3 the camera was added and in 0.3.6 QR code reading was added.

**End of Line**: Preparing for a public demo

## ARIS 0.4 - GLS 2009 Private Release

Includes Mentira branch

**Platform**: Same as above with continued attention to stability. Added a custom HTML based authoring tool, increasing the audience slightly.

**End of Line:** Wanted to complete the transition to a client side tool to improve performance and flexibility.

## ARIS 1.0.x - GLS 2010 Public App Store Release

Includes Bike Box, Love Box, LoVid and other branches

**Platform:** ARIS server now operates with a AMF3 connection to a Flex based authoring tool and a JSON connection to the client. All data is rendered on the two clients.

**New Features:** An alpha release of an easy to use, Flex based authoring tool.

## ARIS 1.1

**Platform:** Same as Above

**New Features**: Localized for Spanish and French. Data caching and network robustness improved greatly. New Nearby object indicator

## ARIS 1.2 - Post 2010 Game Jam Release

**Platform:** Same as Above

**New Features**: Inventory can have quantities. Players can travel to locations on the map without physically moving. Authoring tool Alpha 2 was allowed quests and conversations to be created in addition to using the new requirements for quantities.